\documentstyle[preprint,aps]{revtex}
\begin{document}
\newcommand{\lel}{l_e}

\title{Universal Spectral Correlations in Diffusive Quantum Systems}
\author{Daniel Braun and Gilles Montambaux}
\address{Laboratoire de Physique des Solides, Associ\'e au CNRS,
Universit\'e  Paris--Sud, 91405 Orsay, France}
\date{April 25, 1994}
\maketitle
\begin{abstract}
We have studied numerically several statistical properties of the spectra of
disordered electronic systems under the influence of an Aharonov Bohm flux
$\varphi$, which acts as
 a time--reversal symmetry breaking parameter. The
distribution of curvatures of the single electron energy levels has a
modified Lorentz form with different exponents in the GOE
and GUE regime. It has Gaussian tails in the crossover regime. The typical
curvature is found to vary as $ -E_c\ln (E_c\varphi^2/\Delta)$ ($E_c$ is the
Thouless energy and $\Delta$ the mean level spacing)  and to diverge at zero
flux.  We show that the harmonics
of the variation with  $\varphi$ of single level quantities (current or
curvature) are correlated, in contradiction with the perturbative result.
The single level current correlation function is found to have a logarithmic
behavior at
low flux, in contrast to the pure symmetry cases. The distribution of single
level currents is non--Gaussian in the GOE--GUE transition regime. We
find a
universal relation between $g_d$, the typical slope of
the levels, and $g_c$, the width of the curvature distribution, as was
proposed by Akkermans and Montambaux. We conjecture the validity of our
results for any chaotic quantum system.

\end{abstract}
\newpage
\narrowtext
\section{Introduction}

 Spectral correlations manifest
themselves in many physical properties of mesoscopic systems, for which the
coherence length $l_\varphi$ is much larger than the system size $L$, so
that the electron wavefunctions retain their phase coherence all over the
sample.  It is well known that in the diffusive regime, where
the elastic mean free path
$\lel$ is much smaller than the system size, the low energy
($E\le E_c$) spectral correlations are
 described by Random Matrix
Theory (RMT) \cite{Altshuler86,Efetov83,Mehta}. This description predicts a
level
repulsion and a
correlation of the energy levels over an energy range $E_c=\frac{\hbar
D}{L^2}$, the so called Thouless energy. $D=\frac{v_F\lel}{d}$ is the
diffusion constant of the electrons in the metal, $v_F$ the Fermi velocity
and $d$ the dimensionality of the system. One manifestation of these
spectral correlations is the existence of persistent currents
\cite{Levy,Webb,Mailly93,AGI,Oppen,AE,Schmid} up to temperatures $T$ of the
order
$k_BT\simeq E_c\gg\Delta$, $\Delta$ being the mean level spacing of the
system.

 The ratio $E_c/\Delta$ is a
measure for the conductance $g_c$  of the system in units of $e^2/h$ (we will
henceforth call $g_c$ the ``correlation conductance''). It has been
shown by Thouless \cite{Thouless74} that $E_c$ may be
expressed as the width of the distribution of curvatures
$c=\frac{1}{\Delta}\frac{\partial^2e(\varphi)}{\partial\varphi^2}|_{\varphi=0}$
of a given single
energy level $e(\varphi)$, when the system is closed to a ring and
pierced by an Aharonov--Bohm (AB) flux $\phi=\varphi\phi_0$ ($\phi_0=h/e$ is
the flux quantum). Such an AB--flux may be thought of as introducing
generalized boundary conditions $\psi(x+L)=\psi(x)e^{i2\pi\varphi}$ for the
wavefunctions of the system without AB--flux. Wavefunctions that are
localized on a localization length $\xi\ll L$ (e.g. the typical
wavefunctions of an insulator) do not feel this change in the boundary
conditions, and thus the corresponding energy levels are independent of the
flux ($c=0$). On the other hand, in a metal with large conductance the
wavefunctions extend typically over the whole sample and are thus
sensitive to the boundary conditions, leading to large curvatures $c$. It
has become customary to use the typical curvature $\langle
c^2\rangle^\frac{1}{2}$, i.e. its root mean square with the average
$\langle\ldots\rangle$ taken over different disorder realizations or a
sufficiently large part of the spectrum, as a measure of the width of the
curvature distribution $P(c)$, and  the equation
$g_c=\frac{E_c}{\Delta}=\langle c^2\rangle^\frac{1}{2}$
has become known under the name Thouless--relation \cite{c2}. Little
attention has
been paid so far to the distribution of the curvatures $P(c)$ (or in
general: $P_\varphi(c(\varphi))$, where
$c(\varphi)=\frac{1}{\Delta}\frac{\partial^2e(\varphi)}{\partial \varphi^2}$
is the
``level curvature'' at finite flux) and to the question whether its second
moment exists at all. The only case for which the distribution of curvatures
of energy levels has been examined in some detail is a situation, where a
perturbation parameter $\lambda$ does {\em not} introduce a symmetry change
of the
regarded Hamiltonian $H(\lambda)$ \cite{Rice89,Zakrzewski93,Haake91} (we
will call
these cases the ``pure cases''). In these cases, $H(\lambda)$ is typically
of the form
$H(\lambda)=H_0+\lambda H_1$, where both $H_0$ and $H_1$ belong to the {\em
same} Gaussian ensemble (Gaussian orthogonal ensemble (GOE), Gaussian
unitary ensemble (GUE) or Gaussian Symplectic ensemble (GSE)). At least the
large curvature behavior of $P(c)$ could then be found from Brownian
Motion Models, making use of the fact that the equilibrium distribution of
the eigenvalues is independent of $\lambda$. The obtained result is an
algebraic decay of $P(c)$ for all three pure cases:
$P(c)\propto\frac{1}{c^{2+\beta}}$, with $\beta=1,2,4$ for GOE, GUE,
GSE, respectively \cite{Oppen94}.\\

It is well known that an AB--flux breaks the time--reversal symmetry of the
system and thus introduces a GOE $\rightarrow$ GUE transition, and the
question is, to what extent the results for the pure cases persist for such
a transition. Mathematically the transition may be modelled by

\begin{equation} \label{22RMT}
H(\alpha)=H(S)+i\alpha H(A)\,,
\end{equation}
where $H(S)$ and $H(A)$ are  real symmetric and antisymmetric
matrices of dimension $N$, respectively \cite{French86}. The GOE case
corresponds to $\alpha=0$ , the GUE
case to $\alpha=1$.
The GOE--GUE transition is driven by the combination of parameters
$N\alpha^2$. This RMT description has been used by Dupuis and Montambaux
 to show that the spectral rigidity of the spectrum of
disordered metals is a universal function of the rescaled parameter
$\varphi/\varphi_c$ with $\varphi_c=1/\sqrt{\frac{E_c}{\Delta}}$, which
plays the role of $\sqrt{N}\alpha$ \cite{Gilles92}. Later on supersymmetric
calculations reproduced the RMT results for the spectral correlations in the
transition regime \cite{AEI}.
 Since in this case
 the equilibrium distribution of the eigenvalues changes with the
fictitious time $\alpha$, there has been no successful application of Brownian
Motion Models to determining the curvature distribution $P(c)$. However, a
recent RMT calculation  revealed the following
features \cite{Alex93}:\begin{enumerate}
\item For $\varphi=0$ the large curvature behavior of $P(c)$ is identical to
the one in the pure GOE case,  i.e. $P(c)\propto c^{-3}$ for large
curvatures, which leads to a logarithmic divergence of $\langle c^2\rangle$.
\item For $0<\varphi<\varphi_c$, $P_\varphi(c(\varphi))$ is very different
from  the $P(c)$ for either the pure GOE and the pure GUE case:
It has a Gaussian tail, whereas in the pure GUE case it decays as
$c(\varphi)^{-4}$.

These results have been obtained by exactly solving a $2\times 2$ RMT model
(which should give the correct results for large curvatures), and have been
shown to persist for general $N\times N$ models.\\

Closely related to the second moment of $P(c)$ is the
small flux behavior of certain single level current correlation functions.
Defining the single level current
$i(\varphi)=-\frac{1}{\Delta}\frac{\partial}{\partial\varphi}e(\varphi)$, we
have
$\langle c^2(\varphi)\rangle=\frac{\partial^2}{\partial\varphi\partial
\varphi'}\langle i(\varphi)i(\varphi')\rangle|_{\varphi=\varphi'}$. The
single level
current autocorrelation function
\begin{equation}
C(\varphi_-)=\langle\overline{i(\varphi)i(\varphi+\varphi_-)}\rangle\,,
\end{equation}
where the overbar denotes averaging over one period of the AB--flux, has
been introduced by Szafer and Altshuler \cite{Szafer93}. The
currents $i(\varphi)$ and  $i(\varphi+\varphi_-)$ belong to  the same level.
These authors showed by
diagrammatical perturbation theory and numerical calculations that
$C(\varphi_-)$ exhibits the
universal behavior $C(\varphi_-)= -\frac{1}{\pi^2\varphi_-^2}$ in the
region $\varphi_c\ll\varphi_-\ll\frac{1}{2}$. For $\varphi_-<\varphi_c$ the
perturbation theory breaks
down. In this parameter region, RMT calculations predict  a  logarithmic
small flux behavior for $\langle i(\varphi)i(\varphi')\rangle$ and
$\langle i^2(\varphi)\rangle$, which is in
agreement with the logarithmic divergence of $\langle c^2(\varphi)\rangle$
at zero
flux \cite{Alex93}.\\
In the pure symmetry cases
 results have been obtained by the method of
supersymmetry \cite{Simons93.2}, which  indicate a  quadratic
small flux behavior of $C(\varphi_-)$ in the pure GUE case. It has also
been shown that
 $C(\varphi)/C(0)$ is a universal function of $X=\sqrt{C(0)}\varphi$
for all values of $X$.  This result is reminiscent of the universal
dependence of
the spectral rigidity in terms of $\sqrt{\frac{E_c}{\Delta}}\varphi$ shown
in \cite{Gilles92}.\\

In this work we examine numerically the predictions of \cite{Alex93} and
deduce further consequences. The numerical simulations are performed by
diagonalizing
Anderson tight binding Hamiltonians with diagonal disorder.
The paper is organized as follows: In Section II, we  show that $P(c)$
decays $\propto c^{-3}$ for large $c$ and thus
leads to a logarithmic divergence of $\langle c^2\rangle$. We show that in
the transition regime $P_\varphi(c(\varphi))$ has Gaussian tails and
exhibits a GUE behavior with a $c(\varphi)^{-4}$ decay for large
$c(\varphi)$ in the regime $\varphi\gg\varphi_c$. In section III, we discuss
different current--correlation functions. We find that
 $\langle i^2(\varphi)\rangle$ is proportional to
$\frac{E_c^2}{\Delta^2}\varphi^2\ln(\frac{2\varphi}{\varphi_c})$ for
$\varphi\ll\varphi_c$ and
shows a non--perturbative behavior at any flux. Furthermore,
written in terms of $\varphi/\varphi_c$, the function
$p(\frac{\varphi}{\varphi_c})=\langle
i^2(\varphi)\rangle\varphi_c^2$ becomes  completely universal
for $\varphi\ll 1/2$ with a logarithmic dependence at small flux. A
logarithmic low flux behavior is also found for
 the rescaled correlation function $C(\varphi_-)/C(0)$. By comparing
$\langle i^2 (\varphi)\rangle$ and
$C(\varphi_-)$ we provide direct evidence of the fact that for
$\varphi,\varphi'<\varphi_c$ the product $\langle
i(\varphi)i(\varphi')\rangle$ can not be written in the perturbative
form $f(\varphi+\varphi')-f(\varphi-\varphi')$ in contrast to the pure
symmetry cases. This has deep consequences,
some of which we discuss in section IV, where we reexamine the recently
proposed universal conductance ratio \cite{Gilles92.2} $a= g_c/g_d$, where
$a$ is a universal constant and $g_d$ the dissipative conductance of the
system, which may be expressed as
$g_d=\langle\overline{i^2(\varphi)}\rangle$. We show that
this relation remains valid if for $g_c$ a suitable measure for the width of
the curvature distribution is used, and determine the coefficient $a$
numerically. In section V we are concerned with the distribution of single
level currents $P_\varphi(i(\varphi))$. We confirm a Gaussian behavior of
$P_\varphi(i(\varphi))$ for $\varphi>\varphi_c$ as predicted in
\cite{Simons93.2}, but find deviations for $0<\varphi<\varphi_c$ and small
currents \cite{Alex93}. The
 distribution of
$\overline{i^2(\varphi)}$ is shown to be well described by a log--normal
behavior. Finally, we present our conclusions in section
VI.

\section{Curvature distributions}
\subsection{Curvatures at $\varphi=0$}
As mentioned in the introduction, one expects for $\varphi=0$ an algebraic
decay of $P(c)\propto c^{-3}$ for large curvatures
($c\gg 1$). This
prediction is base on the following arguments: First of all, it is
reasonable to assume that large curvatures
arise from two levels which almost cross, thus from two levels with very
small spacings $s$ at $\varphi=0$ \cite{Rice89,Zakrzewski93,Alex93}.
For such a pair of levels the repulsive interaction with the next levels
can be neglected and they may thus be described by a $2\times 2$ RMT.
Obviously,  the smaller the
spacing $s$ between the two levels, the better this approximation. The
$2\times 2$ RMT can be solved
exactly \cite{Alex93}, but the large curvature behavior is already obtained
from observing that at $\varphi=0$
$c$ varies as $s^{-1}$. This, together with the known small
spacing behavior of the spacing distribution $P(s)$ in GOE, $P(s)\propto
s$, immediately leads to $P(c)\propto c^{-3}$ \cite{Haake91,Shklovskii}.
Whereas the small
curvature behavior of $P(c)$ cannot be obtained from a $2\times 2$ RMT
(for small curvatures the interaction also with the next levels is
important; the exact $2\times 2$ RMT result shows a $c^{-\frac{1}{2}}$
divergence for $|c|<1$, which does not exist), the
evaluation of $P(c)$ from a general $N\times N$ RMT is a difficult problem
\cite{Oppen94}. However, it may be shown rigorously that $P(c)$ at
zero flux must be identical to the $P(c)$ of the pure GOE case. Even though
$\varphi=0$ marks the GOE regime, this is a
priori not
trivial, since the curvatures also probe infinitesimally small fluxes, and the
GOE--GUE transition becomes discontinuous in the limit of infinitely large
matrices.
The only condition for the equality of the two distributions is  that the
matrix elements of the perturbation to which $\varphi$ couples are
independent of those of the unperturbed Hamiltonian. In the transition case
($\varphi$ being an AB--flux), the parameter dependent Hamiltonian may be
written as \cite{French86}  $H(\alpha)=H_1(S)+i\alpha H_2(A)$ with
$\alpha\propto{\varphi\over\varphi_c\sqrt{N}}$. The matrix elements of the
symmetric part
$H_1(S)$ and the antisymmetric part $H_2(A)$ are by definition statistically
independent. $H_1(S)$ belongs to the GOE ensemble, and the matrix elements
of $H_2(A)$ are distributed with the same Gaussian distribution of width
$v$ as those of $H_1(S)$. In the pure GOE case we
have $H(\alpha)=H_1(S)+\alpha H_2(S)$, and both $H_1(S)$ and $H_2(S)$ are
 symmetric random matrices of the GOE ensemble with size $N$. In both
cases, second order perturbation theory in $\alpha H_2$ leads to the exact
relationship

\begin{equation} \label{PTc}
\frac{\partial^2e_n(\alpha)}{\partial\alpha^2}|_{\alpha=0}=\pm 2 \sum_{k\ne
n}^N\frac{H_{2_{kn}}^2}{e_k(0)-e_n(0)}\,,
\end{equation}
where the $+$ and $-$ sign correspond to the transition and pure case,
respectively. $H_{2_{kn}}$ is defined as  $H_{2_{kn}}=\langle
k|H_2|n\rangle$, and $|k\rangle$, $e_k(0)$
are the eigenvectors and eigenvalues of $H_1(S)$, respectively. Writing
$P(c)$ as

\begin{eqnarray}
P(c)&=&\int\left(\prod_{k\ne n}
dH_{2_{kn}}\,P(H_{2_{kn}})\,de_k\right)P(e_1(0)\ldots
e_n(0))\nonumber\\
&&\delta(c\mp 2 \sum_{k\ne n}^N\frac{H_{2_{kn}}^2}{e_k(0)-e_n(0)})\,,
\end{eqnarray}
we notice that the matrix elements $H_{2_{kn}}$ may be integrated out
immediately, if they are independent of  the eigenvalues $e_k(0)$. This
is of course the case if $H_1$ and $H_2$ are two independent random
matrices, regardless whether $H_2$ is symmetric or antisymmetric. The
remaining integral over the $e_k(0)$ is the same in both cases, $P(e_1(0)\ldots
e_n(0))$ being the well known GOE joint distribution function of
eigenvalues. The different sign of the sum in the $\delta$--function is
irrelevant once we consider the curvatures of all levels, by which we obtain
explicitly a symmetric distribution $P(c)=P(-c)$. Thus, $P(c)$ is the
same in both cases.\\

 For the pure GOE case, up to our knowledge  closed
expressions for the full curvature distribution have not been found yet.
However, based on the large
curvature behavior, a formula has been guessed, which
fits  numerically calculated histogramms remarkably well \cite{Zakrzewski93}:

\begin{equation} \label{genlo}
P(c)=\frac{1}{2\gamma}\frac{1}{(1+(\frac{c}{\gamma})^2)^{3/2}}\,.
\end{equation}
The width $\gamma$ of the distribution is proportional to
$\langle i^2(\alpha)\rangle$, which in the pure GOE
case is independent of the perturbation parameter $\lambda$. In agreement
with the above arguments we found exactly the same form of the distribution
for the transition case and we have related the width $\gamma$ to $\langle
\overline{i^2(\varphi)}\rangle$. We
will come back to this point in  section IV.\\

We obtained our numerical data from exactly diagonalizing  tight--binding
Anderson Hamiltonians by a Lanczos procedure. The diagonal elements of the
Hamiltonian are distributed in a box--like distribution from $-w/2$ to
$w/2$. Thus, $w$ is a measure of the disorder, connected to physical
parameters by $\lel\propto\frac{1}{w^2}$ with a prefactor depending on the
dimensionality of the system \cite{Helene91}. The parameter $\varphi$ is
used to introduce
generalized boundary conditions $\psi(x+L,y,z)=e^{i2\pi\varphi}\psi(x,y,z)$
in $x$--direction.   Strictly speaking, $\varphi$ is   only in the case of
quasi--one dimensional samples equivalent to an AB--flux, since for the latter
the introduced phase shift depends in principle on the radial coordinate.
In our numerical calculations it turned out that this radial dependence is
of little importance for the statistical properties of the spectra and we
therefore deal with 2D and 3D samples as well, keeping the phase shift
independent of  the radial coordinate.\\

The precise calculation of the curvatures is a  non--trivial task: In
principle one would like to diagonalize the Hamiltonian for very close flux
values in order to approximate the differentials as well as possible by
differences. The problem is, however, that in each spectrum, especially for
relatively small disorder, large curvatures coexist with very small
curvatures. The latter may not be resolvable anymore for flux values chosen
too close, since the corresponding levels will appear completely flat due to
the finite numerical precision. On the other hand, flux values chosen too
far apart will not allow for a precise calculation of the large curvatures.
We therefore diagonalized the Hamiltonian always for several very small
fluxes as well as for some values leading to flux differences typically an
order of magnitude larger.
 The error of the calculated curvatures were  estimated by fitting the
energy levels to polynomials of different degrees and comparing the results.
 By suitably choosing our flux values we were
 able to obtain a mean relative precision  of the order of $10^{-5}$ for
curvatures absolutely smaller than 90\% of the largest one, $c_{max}$, and
still of the  order of 1\% for curvatures between 0.9$c_{max}$ and
$c_{max}$.\\

Only the central half of the spectrum, where in
three dimensions the density of states is   basically constant, was used.
The curvatures were
rescaled by the mean level spacing and only the distribution of their
absolute value examined, making use of the symmetry of $P(c)$. We used a
large number of disorder realizations (e.g. 100 disorder realizations for 3D
samples up to size 12*12*12 sites, and still 25 realizations for 16*16*16
samples). The obtained distributions were fitted to eqn.~(\ref{genlo}) with an
additional prefactor 2 taking care of the different normalization due to the
absolute
value, and with $\gamma$ as fitting parameter. Fig.\ref{pc0} shows a typical
example. Obviously, eqn.~(\ref{genlo}) describes the observed distribution very
well. The same is true for all other samples examined, regardless of their
dimensionality and their disorder, as long as we stay in the diffusive
regime. The parameters of the  system enter in eqn.~(\ref{genlo}) only in
form of the
width $\gamma$ of the distribution,  and we will see in section IV that this
width can in fact be related to
$\langle \overline{i^2(\varphi)}\rangle$, so that $P(c)$ contains no free
parameter anymore \cite{Zyczkowski94}.
As a  consequence of the distribution eqn.~(\ref{genlo}), its second
moment $\langle c^2\rangle=\int_{-\infty}^\infty P(c)c^2$ diverges
logarithmically
with an upper cutoff $c_{max}$, as was recently found also by \.{Z}yczkowski
et al.~\cite{Zyczkowski94}.

\subsection{$0<\varphi\ll\varphi_c$}

For finite flux the simple relation $c\propto s^{-1}$ does
not hold anymore, but still it is reasonable to assume that the large
curvature behavior of $P_\varphi(c(\varphi))$ can be obtained from
$2\times 2$
RMT. The exact result found in \cite{Alex93} has Gaussian  limiting
behavior, $P_\varphi(c(\varphi))=
a(\varphi)\frac{c(\varphi)}{v}e^{-b(\varphi)\frac{\varphi^2}{\varphi_c^2}\frac{c^2(\varphi)}{v^2}}$, where $a(\varphi)$ and
$b(\varphi)$ are two weakly varying functions of order one in the transition
regime. It is valid for curvatures much larger than
$\frac{v\varphi_c}{\Delta\varphi}$, whereas for smaller curvatures the decay
is
still algebraic as $1/c^3(\varphi)$. For curvatures smaller than $v/\Delta$,
the $2\times 2$ calculation fails to describe the curvature
distribution.
 Since  in the $2\times 2$ RMT model the energy scales $E_c$ and $\Delta$
are not well
separated ($E_c/\Delta\simeq N$, where $N$ is the size of the matrix), one
is not able to decide whether the decay scale $v$ of $P_\varphi(c(\varphi))$
corresponds to the physical parameters $E_c$ or $\Delta$. The numerical
results presented below show that at zero flux  the width
of $P_\varphi(c(\varphi))$ is given by the variance of the single level
current averaged over the flux (thus the Thouless energy). For finite flux
the numerical verification  of the low
flux behavior of the correlation function $C(\varphi,\varphi')$
(eqn.~(\ref{C12RMT}), see Section III)  shows that $v$ has to be identified
with $E_c$.
 As a consequence of the Gaussian tail of the distribution, its
second moment is finite for any  value $\varphi$ of the flux with
$0<\varphi<\frac{1}{2}$.  It
varies like $\langle c^2(\varphi) \rangle^{\frac{1}{2}}\propto
\frac{E_c}{\Delta}\ln\frac{2\varphi}{\varphi_c}$ for $\varphi\ll\varphi_c$
(see also section III).

 In order to detect the Gaussian tail of $P_\varphi(c(\varphi))$
numerically, it is  desirable to take $\varphi$  close  to
$\varphi=\varphi_c$. On the other
hand a large number of curvatures is needed since the probability for large
curvatures decays very rapidly. In Fig.~\ref{pc1} we show the numerically
obtained $P_\varphi(c(\varphi))$ at $\varphi=0.034$ and $\varphi=0.068$
for a 3D  sample (8*8*8 sites, 100 disorder realizations) with $w=7$,
confirming the Gaussian decay.

\subsection{$\varphi_c\ll\varphi\ll\frac{1}{2}$}

For $\varphi>\varphi_c/\sqrt{2}$ but close to $\varphi_c/\sqrt{2}$, a
transition back to an
algebraic decay of $P_\varphi(c(\varphi))$ must take place  \cite{fn1}. For
$\varphi\gg \varphi_c$
 the transition to GUE is completed and the formulas
for the pure GUE case should apply. Thus we expect a $c(\varphi)^{-4}$ decay
of $P_\varphi(c(\varphi))$ for $\varphi\simeq \frac{1}{4}$. This is indeed
what is found numerically, as
shown in Fig.~\ref{pc0} for $\varphi=0.24$. The numerical data are  well
fitted by the
generalized Lorentzian  for the pure GUE regime \cite{Zakrzewski93,Oppen94},
$P_{\frac{1}{4}}(c(1/4))=\frac{2}{\pi\gamma}\frac{1}{(1+(c(\frac{1}{4})/\gamma)^2)^2}$.

\section{Current Correlation Functions}

We now consider the single level current correlation function
\begin{equation}
C(\varphi,\varphi')=\langle
i(\varphi)i(\varphi')\rangle
\end{equation}
 and functions derived from it. It may be related
to spectral density correlations by
$C(\varphi,\varphi')=\frac{\partial^2}{\partial\varphi\partial\varphi'}\langle\delta
N(\varphi)\delta N(\varphi')\rangle$, where $\delta
N(\varphi)=N(\varphi)-\langle N(\varphi)\rangle$ is the fluctuation in the
number of levels in a given energy range $[0,E]$ with $E>E_c$ (contributions
to the flux dependent part of $\delta N(\varphi)$ from energies greater than
$E_c$ are exponentially small \cite{AGI}). This relation is valid for
$\delta N(\varphi)\ll N(\varphi)$, which can be achieved if  the
broadening $\gamma$ of the levels or $E_c\varphi^2$ is much larger than the
mean
level spacing \cite{Szafer93}.
If all the parameters
$\varphi-\varphi'$ and $\varphi+\varphi'$ are larger than $\varphi_c$, we
can use well known results from diagrammatical perturbation theory:
$K(\epsilon,\varphi;\epsilon',\varphi')=\langle
\rho(\epsilon,\varphi)\rho(\epsilon',\varphi')\rangle-\langle\rho(\epsilon,\varphi)\rangle\langle\rho(\epsilon',\varphi')\rangle=K_+(\epsilon,\epsilon';\varphi_+)+K_-(\epsilon,\epsilon';\varphi_-)$,
where $\rho(\epsilon,\varphi)$ is the energy and flux dependent density of
states and

\begin{equation}
K_\pm(\epsilon,\epsilon';\varphi_\pm)=\frac{1}{2\pi^2\hbar^2}\mbox{Re}\sum_{\bf
q}\frac{1}{(D{\bf q}_\pm^2-i(\epsilon-\epsilon')+\gamma)^2}
\end{equation}
is the Cooperon ($K_+$) and Diffuson ($K_-$) contribution
\cite{Altshuler86}. In the case of a thin cylinder, ${\bf q}$ is quantized as
${\bf q}_\pm=(\frac{2\pi}{L_x}(n_x+\varphi_\pm),
\frac{\pi }{L_y}n_y,\frac{\pi}{L_z}n_z)$,  $n_x\in{\cal Z}$, $n_{y,z}\in{\cal
N}$. $L_x$ is the circumference, $L_y \ll L_x$ the thickness and $L_z$ the
height of the cylinder. In terms of $K_\pm(\epsilon,\epsilon';\varphi_\pm)$,
$C(\varphi,\varphi')$ is then given by

\begin{eqnarray}
C(\varphi,\varphi')&=&\int_0^{E}\int_0^{E}d\epsilon
d\epsilon'\nonumber\\
&&\left(\frac{\partial^2}{\partial\varphi_+^2}K_+(\epsilon,\epsilon';\varphi_+)-\frac{\partial^2}{\partial\varphi_-^2}K_-(\epsilon,\epsilon';\varphi_-)\right)\label{C12a}\\
&=&f(\varphi_+)-f(\varphi_-)\label{C12b}\,,
\end{eqnarray}
where $f(\varphi)$ is a periodic and symmetric function of $\varphi$:
$f(\varphi+1)=f(\varphi)$ and $f(-\varphi)=f(\varphi)$. From
$C(\varphi,\varphi')$ the single level current autocorrelation function
\begin{equation}
C(\varphi_-)=\int_0^1C(\varphi,\varphi+\varphi_-) d\varphi
\end{equation}
 can be derived.
The periodicity of $f(\varphi)$ and the fact that $\int_0^1
f(\varphi)\,d\varphi=0$,
yield $C(\varphi_-)=-f(\varphi_-)=-\int_0^{E}\int_0^{E}d\epsilon
d\epsilon'\frac{\partial^2}{\partial\varphi_-^2}K_-(\epsilon,\epsilon';\varphi_-)$.
This function has been analyzed in \cite{Szafer93} and a universal behavior
was found for $\varphi_c\ll\varphi_- \ll\frac{1}{2}$, where $C(\varphi_-)$ is
given by $C(\varphi_-)=-\frac{1}{\pi^2}\frac{1}{\varphi_-^2}$,
independent of any system parameters like disorder, geometry or size of
the sample. It was conjectured that this behavior is characteristic of any
chaotic system. Later on it was shown that $C(\varphi_-)/C(0)$ is  a
universal function of $\sqrt{C(0)}\,\varphi_-$ for all $\varphi_-$
\cite{Simons93.2},
even though no analytical formula could be obtained for the whole parameter
range.\\

For $\varphi_-<\varphi_c$ the diagrammatical Cooperon and Diffuson expansion
of $K(\epsilon;\varphi;\epsilon',\varphi')$ breaks down. An important and
interesting question is, however, whether $C(\varphi,\varphi')$ can still be
written in the form of eqn.~(\ref{C12b}), that is, as a function of the sum
of $\varphi$ and $\varphi'$ minus the same function
of their difference, with a possibly different
function $f(\varphi)$. $2\times 2$ RMT yields the following analytical
result for $C(\varphi,\varphi')$  \cite{Alex93}:
\begin{equation} \label{C12RMT}
C(\varphi,\varphi')\propto
-\frac{E_c^2}{\Delta^2}\varphi\varphi'\ln(\frac{\varphi+\varphi'}{\varphi_c})\mbox{
for } 0\le\varphi,\varphi'\ll\varphi_c\,.
\end{equation}
Differentiating this equation with respect to $\varphi$ and $\varphi'$ and
then putting $\varphi=\varphi'$ leads to $\langle
c^2(\varphi)\rangle\propto -\frac{E_c^2}{\Delta^2}\ln
\frac{2\varphi}{\varphi_c}$ for
$\varphi\ll\varphi_c$, and thus back to a divergence of $\langle
c^2(0)\rangle$.\\
Let us first consider eqn.~(\ref{C12RMT}) for $\varphi=\varphi'$, for which it
yields

\begin{equation} \label{fit}
\langle
i^2(\varphi)\rangle=-P\frac{\varphi^2}{\varphi_c^4}\ln(\frac{2\varphi}{\varphi_c})
\end{equation}
with some constant prefactor $P$.
Thus, we can obtain $\varphi_c$ (and the
prefactor $P$) from fitting eqn.~(\ref{fit}) to the numerically calculated
function $\langle
i^2(\varphi)\rangle$. We calculated this function for many
quasi--one dimensional, two--, and three--dimensional samples in the
diffusive regime for different values of the disorder and from
many disorder realizations. Fig.~(\ref{i2}) shows a typical example obtained
from a 2D sample with 28*27 sites and $w=1.9$.
Eqn.~(\ref{fit}) fits the numerical data very well up to $\varphi$
of the order of $\varphi_c$. It fits much better than a simple power
expansion in  $\varphi^2$ with two parameters, which would be the
perturbational
result extended to $\varphi<\varphi_c$. Also, the disorder dependence of the
fit parameters is as expected: Since
$\varphi_c=\frac{1}{\sqrt{E_c/\Delta}}$, $E_c=\frac{\hbar v_F\lel}{dL_x^2}$
and $\lel\propto\frac{1}{w^2}$ \cite{Helene91}, we have $\varphi_c\propto w$.
All the disorder dependence of the prefactor should be contained in the
factor $1/\varphi_c^4$, so that $P$ is independent of disorder.
 Both
relations are  well observed numerically, and, in fact, $P$ is found to be
the same for all systems examined (within numerical precision). This leads
us to the following natural
rescaling of $\langle i^2(\varphi)\rangle$:
\begin{equation} \label{C11res}
p(x)= \langle i^2(\varphi)\rangle\varphi_c^2=-Px^2\ln2x\,.
\end{equation}
Thus, $p(x)$ is a
 universal function of $x=\frac{\varphi}{\varphi_c}$,
independent of any system specifications, as long as we stay in the
diffusive regime. This statement holds so far for $\varphi\le\varphi_c$.
However, since it is found numerically that $\langle
i^2(\varphi)\rangle$ is basically independent of $\varphi$ for $\varphi_c\le
\varphi\le\frac{1}{4}$ and symmetric with respect to $\varphi=\frac{1}{4}$,
one is led to the conclusion that $p(x)$ is a universal
function of all $\varphi/\varphi_c$, as long as the zero--mode approximation
of $\langle i^2(\varphi)\rangle$ holds. For larger  flux values, the
periodicity constraint
 $\langle
i^2(\varphi)\rangle=\langle i^2(\varphi+\frac{1}{2})\rangle$ makes higher
modes important. By the
rescaling the period of this function becomes $\frac{1}{2\varphi_c}$ and
thus non--universal.\\
Due to the RMT origin of eqn.~(\ref{C11res}), we conjecture
the universality of $p(x)$ more generally for any chaotic
system. Note that due to the universality of $p(x)$ up to values $x>
1$  two different
ways of rescaling are possible in principle: The first one, described above,
starts from
determining $\varphi_c$ from the low flux behavior of
$\langle i^2(\varphi)\rangle$ and then defining
$\langle i^2(\varphi)\rangle\varphi_c^2$. The
second starts from rescaling the amplitude $\psi=\langle
i^2(\varphi=\frac{1}{4})\rangle$ and then considering
$\langle i^2(\varphi)\rangle/\psi$ as a function of $\varphi/\sqrt{\psi}$.
Both ways are equivalent --- the
rescaling applies to amplitude and argument of $\langle
i^2(\varphi)\rangle$. We are thus lead to the conclusion that
\begin{equation} \label{cphic}
\varphi_c\sqrt{\psi}=const\,.
\end{equation}
The amplitude $\psi$ has been used before \cite{Zyczkowski94} as a measure of
the width of the curvature distribution at zero flux, thus of $E_c$.
In
Fig.~\ref{i2res} we demonstrate the universality of $p(x)$ numerically for
several
systems. Using the second method, which is numerically simpler, the
amplitude $\psi$ is rescaled to 1. Renormalizing by fitting eqn.~(\ref{C11res})
to the low flux behavior leads to the numerical coefficient
$P=17.78\pm 2.40$, and the constant in eqn.~(\ref{cphic}) is given by $0.96\pm
0.12$. The error margins are simple
standard deviations calculated from seven samples.

Having thus gained confidence in the validity of eqn.~(\ref{C12RMT}) for
$\varphi=\varphi'$ and having determined the parameters $P$ and $\varphi_c$,
we now
consider the equation for $\varphi\ne\varphi'$ for the same systems. In Fig.
\ref{C12} we have plotted the numerically obtained $C(\varphi,\varphi')$ for
given $\varphi'$ as a function of $\varphi$ together with the rhs of
eqn.~(\ref{C12RMT}). The agreement is satisfactory up to $\varphi$ of the
order of $\varphi_c$. Note that $C(\varphi,\varphi')$ is not fitted to the
data for $\varphi\ne\varphi'$, but is completely determined by $P$ and
$\varphi_c$
obtained from $\langle i^2 (\varphi)\rangle$.
 The  arguments concerning rescaling
and universality given above for $\langle i^2(\varphi)\rangle$ for
$0\le\varphi\ll\varphi_c$ apply in this parameter range to
$\langle i(\varphi)i(\varphi')\rangle$ as well.\\

We now come back to the question of writing $C(\varphi,\varphi')$ in the
form of eqn.~(\ref{C12b}) for $\varphi$ or $\varphi'$
smaller than $\varphi_c$.
 This question is equivalent
to the question of the decoupling of the harmonics of $C(\varphi,\varphi')$:
Let us expand $i(\varphi)$ (the current attributed to
some given level) in a Fourier series: $i(\varphi)=\sum_{p=1}^\infty
i_p\sin(2\pi p\varphi)$, which is the most general form compatible with the
symmetry and periodicity requirements of $i(\varphi)$. Then we have
$C(\varphi,\varphi')=\sum_{p,q}\langle i_pi_q\rangle\sin(2\pi
 p\varphi)\sin(2\pi q\varphi')$. It may be shown that the form (\ref{C12b}) of
$C(\varphi,\varphi')$ for all $\varphi,\varphi'$ is equivalent to $\langle
i_pi_q\rangle=\langle i_p^2\rangle\delta_{p,q}$, that is that the harmonics
of $\langle i(\varphi)i(\varphi')\rangle$ decouple.

Having shown that $C(\varphi,\varphi')$ is well
described by eqn.~(\ref{C12RMT}), it is obvious that  $C(\varphi,\varphi')$
can {\em not} be
written in the form (\ref{C12b}) for $\varphi,\varphi'<\varphi_c$ and that
therefore the harmonics of $C(\varphi,\varphi')$ do {\em not} decouple.
There is a  second way of showing this,  which is even more direct: Let us
{\em assume} that
$C(\varphi,\varphi')$ can be written for all $\varphi,\varphi'$ with
$0\le\varphi,\varphi'\le 1$ in the form
of eqn.~(\ref{C12b}). Then $\langle i^2 (\varphi)\rangle$ is given by
$\langle i^2 (\varphi)\rangle=f(2\varphi)-f(0)$ and $C(\varphi_-)$ by
$C(\varphi_-)=-f(\varphi_-)$. Thus we would have the following connection
between  $\langle i^2 (\varphi)\rangle$ and $C(\varphi_-)$:

\begin{equation} \label{connect}
\langle i^2 (\varphi)\rangle=C(0)-C(2\varphi)\,,
\end{equation}
i.e. $\langle i^2 (\varphi)\rangle$ and $C(\varphi_-)$ would be essentially the
same functions up to a stretching by a factor 2 of the $\varphi$--axis, a
shift and a change of sign. However, eqn.~(\ref{connect}) is not at all
observed numerically. In Fig.~\ref{cmp} we show the two functions,
$\langle i^2 (\varphi)\rangle$ calculated directly, and $C(0)-C(2\varphi)$
obtained from $C(\varphi_-)$ via eqn.~(\ref{connect}) for the same samples.
Whereas $\langle i^2 (\varphi)\rangle$ is more or less constant for
$\varphi>\varphi_c$, $C(0)-C(2\varphi)$ reaches a maximum for
$\varphi\simeq\varphi_c$ and then decays again until $\varphi=1/4$. This
decay corresponds to the universal $-\frac{1}{\pi^2\varphi_-^2}$ behavior of
$C(\varphi_-)$ for $\varphi_c\ll\varphi_-<\frac{1}{2}$. Thus,
eqn.~(\ref{connect}) does {\em not} hold and consequently the assumption
that
$C(\varphi,\varphi')$ may be written in the form of eqn.~(\ref{C12b}) for all
$\varphi,\varphi'$ is wrong. \\

Since we know from perturbation theory that $C(\varphi,\varphi')$ has the
form of  eqn.~(\ref{C12b}) for $|\varphi_-|,|\varphi_+|>\varphi_c$,
$C(\varphi,\varphi')$ must differ from the form in eqn.~(\ref{C12b}) for
some $|\varphi_-|<\varphi_c$ or $|\varphi_+|<\varphi_c$. The latter
condition leads for $0\le\varphi,\varphi'\le 1$ to
$\varphi,\varphi'<\varphi_c$, which implies again $|\varphi_-|<\varphi_c$.
Thus, the most restrictive statement we can make
from the above proof is that $C(\varphi,\varphi')$ for
$0\le\varphi,\varphi'\le 1$ is not of the form of eqn.~(\ref{C12b}) for {\em
some} (not neccessarily all!) $\varphi,\varphi'$ for which
$|\varphi_-|<\varphi_c$. The numerical data shown above for
$\langle i(\varphi)i(\varphi')\rangle$ strongly suggest
that this is the case  for $0<\varphi,\varphi'<\varphi_c$. Finally we note
that $\langle i^2(\varphi)\rangle$ is
{\em non--perturbative for all flux values}, since $\varphi_-=0$ for all
$\varphi$.
\\

We now investigate more closely the low flux behavior of $C(\varphi_-)$. As
mentioned, $C(\varphi_-)$ can not be obtained from perturbation theory for
$|\varphi_-|<\varphi_c$, and from SUSY so far only in the pure GUE regime
\cite{Simons93.2}.
Comparing $\langle i^2(\varphi)\rangle$ and
$C(0)-C(2\varphi)$ in Fig.~\ref{cmp}, we find that their low flux
behavior looks very similar, even though not up to the $\varphi_c$ defined
by $\langle i^2(\varphi)\rangle$ via eqn.~(\ref{C11res}).
The latter restriction must not surprise: Note that the area under the two
curves must be the same: $\int_0^1 i^2(\varphi)\,d\varphi=C(0)$ and
$\int_0^1  (C(0)-C(2\varphi))\,d\varphi=C(0)$, since $\int_0^1
C(\varphi)\,d\varphi=0$ \cite{Szafer93}. But since the two functions are
different for
$\varphi>\varphi_c$, they also have to differ for some  $\varphi<\varphi_c$.
Nevertheless, the similar behavior for $\varphi\ll\varphi_c$ makes it
reasonable to fit $C(\varphi)$ to $C(0)+P'\frac{\varphi^2}{\varphi_c'^4}\ln
(\varphi/\varphi_c')$ with parameters $P'$, $\varphi_c'$ possibly different
from  $P$ and $\varphi_c$. This fit works indeed
very well, and, like for $\langle
i^2(\varphi)\rangle$, much better than a power expansion to  the fourth
order in $\varphi$
 (see Fig.\ref{Cofphi}).
The latter corresponds to
the  SUSY result
 for the pure GUE case, $C(\varphi)/C(0)\simeq
1-2\pi^2X^2$ with $X=\varphi\sqrt{C(0)}$ \cite{Simons93.2}, and also to the
low--flux expansion of a formula recently proposed   by Delande and Zakrzewski
in the pure GOE regime \cite{Delande94}. Therefore  the parts
in the integral over flux in the definition of $C(\varphi)$ arising from the
transition regime contribute crucially and lead to a behavior significantly
different from the one in the pure symmetry  cases, even though the transition
region is typically small compared to the region of pure GUE behavior
\cite{Simons93.2}.
The coefficients $\varphi_c'$ and
$P'$ depend in the same way as $\varphi_c$ and $P$ on the disorder
($\varphi_c'\propto w$ and $P'$ independent of $w$), and we obtain for the
rescaled
function $C(\varphi)/C(0)$ the universal low--flux
behavior
\begin{equation}
C(\varphi)/C(0)\simeq 1+P'x'^2\ln x'
\end{equation}
 with $P'=6.67\pm 0.14$ and
$x'=\varphi/\varphi_c'$. The error margin is the simple standard deviation
calculated from four macroscopically different samples with up to 100
disorder realizations.

\begin{center}
{\bf IV. THE UNIVERSAL CONDUCTANCE RATIO $ a=g_c/g_d$}
\end{center}

In \cite{Gilles92.2} a different measure $g_d$ of the average conductance of
an ensemble of macroscopically identical mesoscopic samples in the
diffusive regime was
introduced:
$g_d=\langle\overline{i^2(\varphi)}\rangle$, where
$i(\varphi)$ is the single level current of a given level close to the
Fermi level, and the overbar denotes an average over $\varphi$ taken over one
period. It was shown that this conductance is the average dissipative
conductance of the ensemble of  samples, equivalent to the one derived
from the Kubo--formula in the limit of vanishing frequency. Similar formulas
have been used in the literature to express the width of curvature
distributions (see
e.g.\cite{Zakrzewski93,Zyczkowski94}) and have also been applied to cases
where $\varphi$ represents not an AB--flux but an arbitrary parameter in the
system (like the system size, or an electric field) \cite{Simons93.2}. Writing
the conductance as
$g_d=\langle\overline{i^2(\varphi)}\rangle$ is another
way of expressing the conductance as the sensitivity of the system's
spectrum to a change of the boundary conditions. Using only two assumptions,
namely $i)$ $\langle
i(\varphi)i(\varphi')\rangle=\frac{\partial^2}{\partial\varphi\partial\varphi'}\langle\delta
N(\varphi)\delta N(\varphi')\rangle$  and $ii)$  $\langle\delta
N(\varphi)\delta
N(\varphi')\rangle=F(\varphi+\varphi')+F(\varphi-\varphi')$ for all $0\le
\varphi,\varphi'\le 1$ (with $F(\varphi)=F(\varphi+1)=F(-\varphi)$) it was
also shown that the ratio $a=\frac{g_c}{g_d}=\frac{\langle
c^2\rangle^\frac{1}{2}}{\langle\overline{i^2(\varphi)}\rangle}$
should be a universal one. Making the above two assumptions, its value
follows from RMT to be $a=\pi\sqrt{12}$. Having found that $\langle
c^2\rangle$ diverges , we now know that this relation cannot hold in the
above form. This is due to the fact that
$\langle
i(\varphi)i(\varphi')\rangle$ cannot be written as
$f(\varphi+\varphi')-f(\varphi-\varphi')$ for all $0\le\varphi,\varphi'\le
1$, as was shown in the previous section. Thus, if the first assumptions
holds, the second one is violated. However, if we use a correct measure for
$g_c$, like e.g.~$g_c=\langle|c|\rangle$, the qualitative arguments given in
Ref.~\cite{Gilles92.2} should still be correct and $g_c/g_d$ {\em
should} be a universal ratio. This universality is based on the fact that
the GOE$\rightarrow$GUE transition is driven by the unique parameter
$\varphi/\varphi_c$. Thus, different measures of the spectra's sensitivity
to changes of the boundary conditions should be equivalent. Unfortunately,
we did not succeed in giving a RMT prediction of the ratio $\frac{\langle
|c|\rangle}{\langle\overline{i^2(\varphi)}\rangle}$, but we
checked the proportionality between the two quantities numerically. The
linear relation between $\langle|c|\rangle$ and
$\langle\overline{i^2(\varphi)}\rangle$ is well observed
in the diffusive regime in the form $\langle|c|\rangle=b+a
\langle\overline{i^2(\varphi)}\rangle$ with
$b=13.4$, $a=6.7$ for different sample geometries and values
of disorder (see Fig.~\ref{prop}). We estimate the error margin for the slope
to be of the order of 10\%. Similar relations have been reported
before, e.g.~in \cite{Zakrzewski93} for the pure symmetry cases without
taking an average over the perturbation parameter and in \cite{Zyczkowski94},
where $\langle|c|\rangle\propto\langle
i^2(\varphi=\frac{1}{4})\rangle$ was found. One consequence of this result
is that the rescaled curvature distribution $g_dP(c)$ should be a
universal function of $c/ g_d$.
In Fig.~\ref{pcuni} we demonstrate this universality numerically.

\begin{center}
{\bf V. CURRENT DISTRIBUTIONS\\}
\end{center}

In this section we examine the distributions $P_\varphi(i(\varphi))$ and
$P(\overline{i^2(\varphi)})$. The former has been investigated in
\cite{Simons93.2} in the pure GUE case for a perturbation by an AB--flux or
by an external potential. Employing SUSY methods all the moments of the
current distribution were obtained and found to be in agreement with a pure
Gaussian distribution for all currents at fixed flux. In \cite{Alex93} the
question of the crossover behavior of $P_\varphi((i(\varphi))$ from GOE to
GUE was addressed, and the following asymptotic behavior was found: An
algebraic decay as $1/i^3(\varphi)$ for currents smaller than $v$ and a
Gaussian decay on the scale $v$ for larger currents. Concerning the
identification of $v$ with physical parameters, the arguments  given for
the discussion of $P_\varphi(c(\varphi))$ apply. Thus, the scale of the
decay is $E_c$ and independent of $\varphi$. \\

The current distribution function $P_\varphi(i(\varphi))$ is of direct
physical interest, since it has been shown \cite{Bertram93} that
$\langle i^2(\varphi)\rangle$ is proportional to the diagonal AC
conductance of a mesoscopic ring pierced by an AB--flux. We may therefore
expect that $P_\varphi(i^2(\varphi))$ gives the corresponding conductance
fluctuations. Knowing $P_\varphi(i(\varphi))$, we obtain immediately
$P_\varphi(i^2(\varphi))=\frac{1}{\sqrt{i^2(\varphi)}}P_\varphi(\sqrt{i^2(\varphi)})$.
In Fig.~\ref{Pofi} we show the numerically obtained current distibution for
several values of the flux. For $\varphi\gg\varphi_c$,
$P_\varphi(i(\varphi))$ is a Gaussian distribution for all values of the
current. The width of the distribution is for
$\varphi_c\le\varphi\le\frac{1}{2}-\varphi_c$  basically independent of
$\varphi$, in agreement with the almost constant value of
$\langle i^2(\varphi)\rangle$ in this parameter range,
found in section III. For $0<\varphi<\varphi_c$, only the tail of the
distribution is Gaussian,  in agreement with the RMT prediction.
\\

We now consider the distribution of $\overline{i^2(\varphi)}$,
$P(\overline{i^2(\varphi)})$. With the result of \cite{Gilles92.2} that
$g_d=\langle\overline{i^2(\varphi)}\rangle$ is the average dissipative
conductance of an ensemble of macroscopically identical samples one might
suppose that $\overline{i^2(\varphi)}$ should be distributed according to
the formulas of universal conductance fluctuations (UCF). Thus,
$P(\overline{i^2(\varphi)})$ is expected to be a Gaussian with a constant
width of order one and
centered around $ \langle\overline{i^2(\varphi)}\rangle$. Surprisingly
enough, this is not at all observed numerically.
Rather, the data are well fitted by a log--normal distribution  with a
non--universal width $s$ (see
Fig.~\ref{Pofi2}), which in the examined disorder region is a linear
function of the disorder $w$: $s\simeq 0.20+0.11 w$.
One possible reason for this non--universal behavior might be that we are
dealing with a closed system. Deviations from Gaussian distributions of
conductances have been reported for samples weakly coupled to leads
\cite{Serota87,Prigodin94}.

\begin{center}
{\bf VI. CONCLUSIONS\\}
\end{center}

We examined in this paper different measures of the sensitivity of the
spectra of diffusive electronic systems to a change of the boundary
conditions, as it
is relevant for a crossover from GOE to GUE. We have
proved analytically and have confirmed
numerically that the single level curvature distribution function is for
zero flux identically the same as in the pure GOE case. For large curvatures
it decays $\propto c^{-3}$ and leads to a logarithmic divergence of its
second moment. From this follows the neccessity to express the Thouless
energy  $E_c$
in terms of the first moment of the curvature distribution, and we showed
that in this case the ratio $E_c/(\Delta
\langle\overline{i^2(\varphi)}\rangle)$ is a universal one. For finite flux
smaller than the critical flux $\varphi_c$, which marks the crossover to the
GUE regime, the curvature distribution was confirmed to be Gaussian, whereas
for $\varphi\gg\varphi_c$ the formulas for the pure GUE case apply and we
get a $c^{-4}$ decay for large curvatures. The distribution of single
level currents was shown to be Gaussian for large currents and all
$\varphi$, whereas for
$\varphi<\varphi_c$ and small currents it decays algebraically.
Finally, we confirmed numerically that the single level current correlation
function
$C(\varphi,\varphi')=\langle i(\varphi)i(\varphi')\rangle$ shows a
logarithmic small flux behavior and cannot be written as
$f(\varphi+\varphi')-f(\varphi-\varphi')$ for $\varphi,\varphi'<\varphi_c$.
$C(\varphi,\varphi')$ becomes universal after appropriate rescaling. The
corresponding function for $\varphi=\varphi'$
shows a non--perturbative behavior for all  $\varphi$.
A logarithmic small flux behavior was also found for the
single level current autocorrelation function. Thus, it seems that a
logarithmic low--flux behavior is a general property of a whole class of
correlation functions envolving flux--derivatives of a given energy level:
$\langle i^2(\varphi)\rangle\varphi_c^2\propto -x^2\ln 2x$,
$C(\varphi)/C(0)-1\propto -x'^2\ln x'$, $\langle
c^2(\varphi)\rangle\propto
-\varphi_c^{-4}\ln\left(\frac{2\varphi}{\varphi_c}\right)$.
Due to the RMT origin of our results, we conjecture their validity for any
chaotic quantum system.

\acknowledgments
We have benefitted from many useful discussions with E.~Akkermans,
H.~Bouchiat,Y.~Gefen, A.~Kamenev and B.~Shklovski\u{\i}. Numerical
calculations were performed on the CRAY computer of the Centre de Calcul
Vectoriel pour la Recherche (CCVR) at Palaiseau and IDRIS (Orsay). This work
was supported in part by
the EC Science program  under grant number $SSC_-CT90_-0020$.\\

{\em Note added:} While completing this manuscript we received a preprint
where the logarithmic low flux
 behavior of the correlation function was found by a
supersymmetric calculation \cite{Taniguchi94}.

\begin{figure}
\caption{The numerically calculated curvature distributions at $\varphi=0$
(empty circles) and
$\varphi=0.24$ (full circles) for 100 8*8*8 samples with $w=5$. The full
lines are
fits to the generalized Lorentzians known from the pure GOE and GUE case
with the widths as fitting parameters. The curvature defined
here is a dimensionless quantity.
  }\label{pc0}
\end{figure}

\begin{figure}
\caption{The curvature distribution at fluxes $\varphi$ between 0 and
$\varphi_c$. The log--log plot of the big graph shows the deviations from
the pure GOE behavior.
The large curvature behavior is Gaussian, corresponding to
straight lines in the inset of the plot. Empty circles correspond to
$\varphi=0.034$, full  circles to $\varphi=0.068$.
  }\label{pc1}
\end{figure}

\begin{figure}
\caption{The low flux behavior of $\langle
i^2(\varphi)\rangle$.  The full line is the fit to eqn.~(\protect\ref{fit})
, the dotted line the fit to $A\varphi^2+B\varphi^4$.}\label{i2}
\end{figure}

\begin{figure}
\caption{The rescaled function $\langle i^2(\varphi)\rangle/\langle
i^2(1/4)\rangle$ as a function of $\varphi\protect\sqrt{\langle
i^2(1/4)\rangle}$
for several two and three dimensional samples.}
\label{i2res}
\end{figure}

\begin{figure}
\caption{Comparison of $\langle
i^2(\varphi)\rangle$ (full circles) with $C(0)-C(2\varphi)$ (empty circles).
Whereas the first is
basically constant for $\varphi>\varphi_c$, the latter reaches a maximum at
$\varphi\simeq\varphi_c$ and then decays again.
  }\label{cmp}
\end{figure}

\begin{figure}
\caption{The function $\langle
i(\varphi)i(\varphi')\rangle$ for $\varphi\ne\varphi'$ and fixed
$\varphi'=0.045\simeq 0.18\,\varphi_c$ (empty circles) for a 2D system
(28*27 sites, $w=1.9$). The solid and dotted lines are the
$2\times 2$ RMT predictions, the two free
parameters being obtained from $\langle
i(\varphi)^2\rangle$ (full circles)}\label{C12}
\end{figure}

\begin{figure}
\caption{The low flux behavior of $C(\varphi)/C(0)$ (full circles) is well
described by $1+P'\frac{\varphi^2}{\varphi_c'^2}\ln(\varphi/\varphi_c')$
(full line). The
dotted  line  represents the pure GUE behavior:
$C(\varphi)/C(0)=1-2\pi^2C(0)\varphi^2$.
  }\label{Cofphi}
\end{figure}

\begin{figure}
\caption{The linear relation between the two conductances
$g_c=\langle |c|\rangle$ and
$g_d=\langle \overline{i^2(\varphi)}\rangle$ for several
two  and three dimensional samples (crosses and diamonds, respectively). All
data points are calculated from 50 or
100 disorder realizations.
  }\label{prop}
\end{figure}

\begin{figure}
\caption{The universal curvature distribution at  zero flux for several
systems (8*8*8 sites with $w=4,5,6$) after rescaling.
  }\label{pcuni}
\end{figure}

\begin{figure}
\caption{The current distribution $P_\varphi(i(\varphi))$ for a 3D sample
(8*8*8 sites and $w=5$) for several
values of $\varphi$ in the range $0<\varphi<\varphi_c$ (crosses:
$\varphi=0.017$, stars: $\varphi=0.051$, empty circles: $\varphi=0.086$) and
for
$\varphi=\frac{1}{4}$ (full circles). In the latter case the distribution is
Gaussian
(corresponding to a straight line in the above plot) for
all currents, whereas for small fluxes only the tails of the distributions
are Gaussian. The slope for large currents is the same for all values of the
flux.
  }\label{Pofi}
\end{figure}

\begin{figure}
\caption{The distributions of $\overline{i^2(\varphi)}$
for several 8*8*8 samples (100 disorder realizations each) with the disorder
parameter $w$ varying  between $w=4$ (crosses) and $w=7$ (full circles). The
distributions are
well fitted by a log--normal distributions (solid lines).
  }\label{Pofi2}
\end{figure}


\begin{references}
\bibitem{fn1} Note that our definition of $\varphi_c$ differs by a factor
$\sqrt{2}$ from the one given in Ref.~\cite{Alex93}.
\end{references}
\end{document}